\documentclass[12pt]{article}
\usepackage{times}
\usepackage{amsmath}
\usepackage{amssymb}
\pdfoutput=1
\usepackage{graphicx}         
\usepackage{epstopdf}         
\usepackage{ITC}
\usepackage{ifpdf}
\ifpdf
\else
\fi
\title{\Large \vspace{-0.5in}
\MakeUppercase{Polarization diversity and equalization of frequency selective channels in telemetry environment for 16APSK
 }}
\author{FARAH ARABIAN\\
    \normalsize Brigham Young University \\
    \normalsize Department of Electrical and Computer Engineering\\
    \normalsize Provo, UT, 84602\\
    \normalsize farah.arabian@gmail.com\\[6pt]
    Faculty Advisor: \\
    Dr. Michael Rice
}
\date{}
\begin{document}
\maketitle 
\section{\MakeUppercase{Abstract}}
\noindent
 Providing RHCP and LHCP outputs from the antenna’s vertical (V) and horizontal (H) dipoles in the resonant cavity within the antenna feeds is the current practice of ground-based station receivers in aeronautical telemetry. 
The equalizers on the market, operate on either LHCP or RHCP alone, or a combined signal created by co-phasing and adding the RHCP and LHCP outputs.
In this paper, we show how to optimally combine the V and H dipole outputs and demonstrate that an equalizer operating on this optimally-combined signal outperforms an equalizer operating on the RHCP, LHCP, or the combined signals. Finally, we show how to optimally combine the RHCP and LHCP outputs for equalization, where this optimal combination performs as good as the optimally combined V and H signals.  
\section{\MakeUppercase{Introduction}}
\noindent
Multi-path between transmitter and receiver leads to inter-symbol-interference (ISI), which affects the performance of the transmission. Multi-path interference can be mitigated using different diversity techniques such as polarization diversity. Right-hand-circular polarization (RHCP) and left-hand-circular polarization (LHCP) are common polarization diversity techniques used in the telemetry applications instead of the linear polarizations such as vertical and horizontal polarizations; the reason is if the airplane goes through different rotations it is hard for the transmitter antenna to get aligned with the vertical or horizontal elements of the receiver antenna.

Current receivers in the telemetry applications synthesize their inputs to come up with RHCP and LHCP signals.
In this paper we will explore the interaction of the receiver antenna in a telemetry environment, in other words,
we will explore if one could access to the receiver's antenna elements signals directly, it means before synthesizing the inputs to get the circular polarizations components then how one could manipulate them to get a better system performance in bit-error-rate (BER) point of view.
\section{\MakeUppercase{The system model and analysis}}
\label{sec:Body}
	\begin{figure}
	\centering
	\includegraphics[width=4.7in]{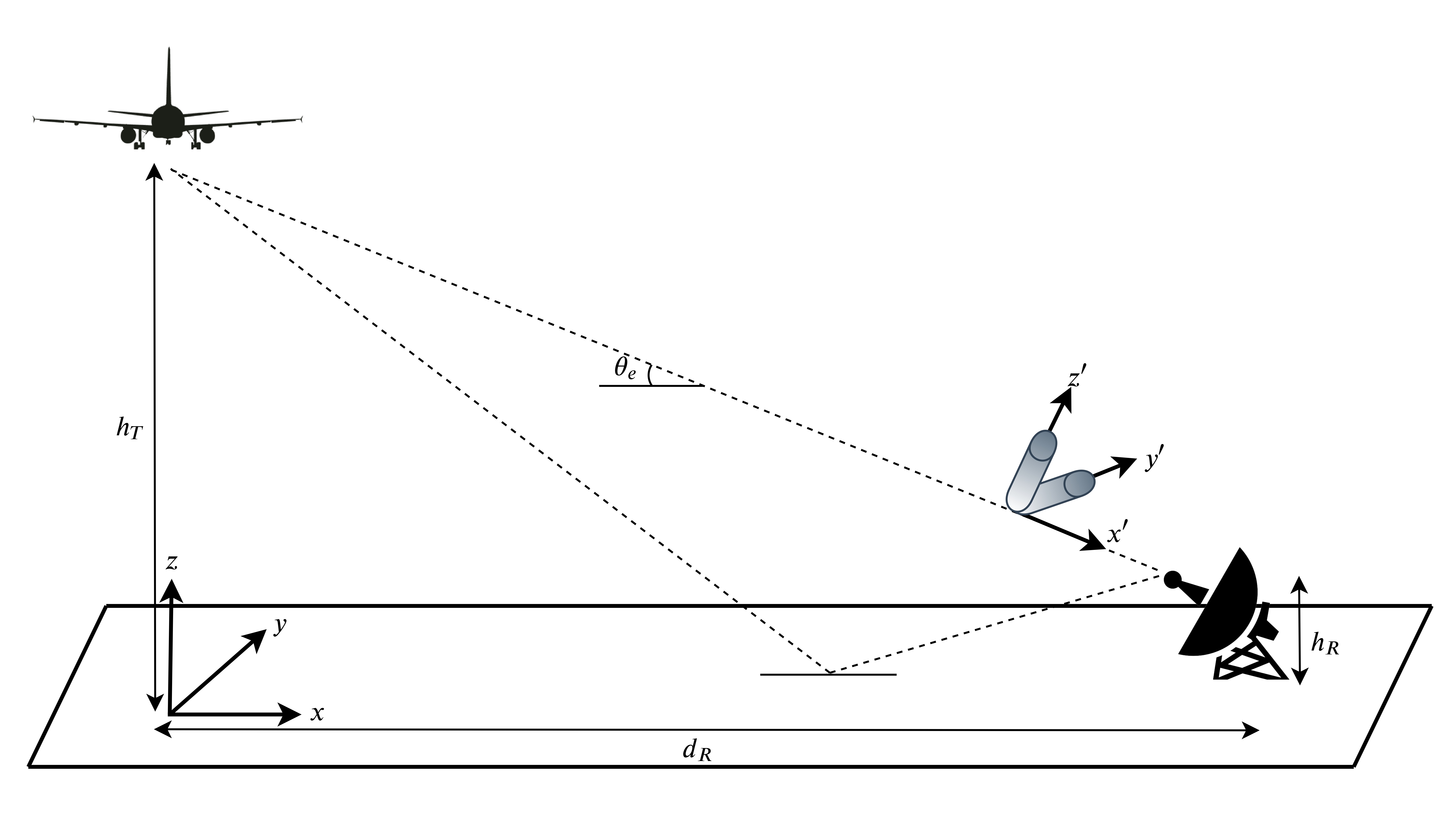}
	\caption{Geometry of two paths radio propagation in the telemetry application.}
	\label{fig:scenario}
\end{figure}
The scenario used in this paper is indicated in Figure~\ref{fig:scenario}, two coordinate systems are used, $(x,y,z)$ coordinate system that is centered right below the transmitter antenna on the ground and the $(x',y',z')$ coordinate system, which is centered at the receiver antenna. The latter coordinate system is related to the former one by a translation along the x-axis and by a rotation over the y-axis by the antenna elevation angle $\theta_{e}$, defined as
\begin{equation}
\theta_e= \tan^{-1}\left(\frac{h_T-h_R}{d_R}\right),
\end{equation}
where $h_T$ and $h_R$ are the height of the transmitter and receiver antennas (in meter) from the ground respectively, and $d_R$ is the ground distance between the transmitter and receiver antennas.

The transmitted signal reaches the receiver through one line-of-sight (LOS) path and one reflected path (NLOS) due to a ground bounce~\cite{MDR_2004}. Both LOS and the NLOS paths are located in the $x-z$ plane while the airplane is in the level flight, means when the wings of the airplane are parallel with the $x-y$ plane.
The transmitter antenna is a vertical dipole mounted on the bottom of the airplane, which is aligned with the $z$ axis only for the level flight and the receiver is a parabolic reflector antenna. There is a resonant cavity at the focal point of the receiver antenna, the resonant cavity is equipped with a cross dipole with one element in the $y'$ direction and the other in the $z'$ direction in the $(x',y',z')$ coordinate system, which means any portion of the received electrical field in the $x'$ direction will be missed in the receiver.

 The receiver antenna only see the $z'$ component of the electrical field in the level flight, but as the airplane goes through the yaw, pitch and roll rotations, which are translations along the x-axis and rotations over $z$, $x$, and $y$ axes respectively in the $(x,y,z)$ coordinate system, both antenna elements in the receiver will detect a portion of the received electrical field.

\begin{figure}
	\centering
	\includegraphics[width=7.1in]{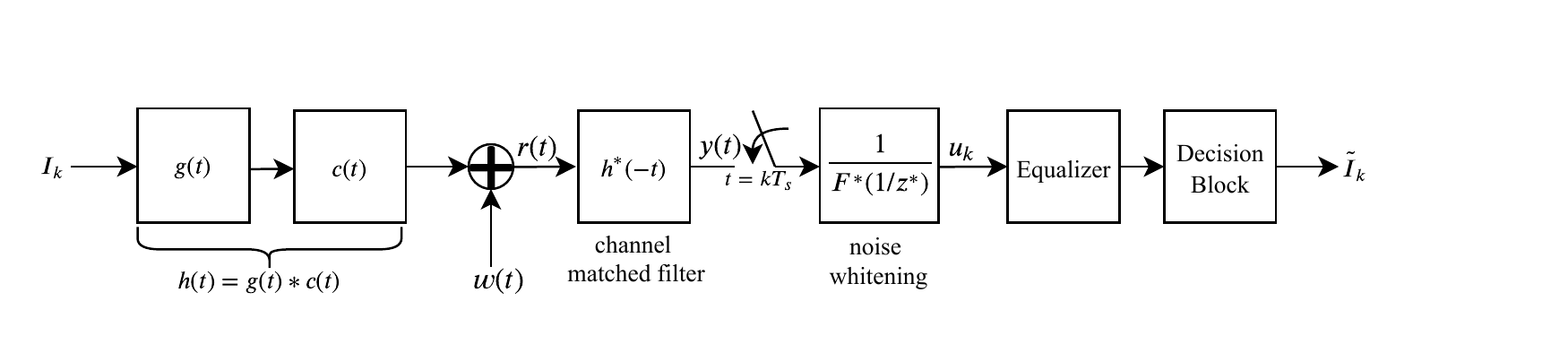}	
	\caption{The Forney observation model.}
	\label{FOM}
\end{figure}

\begin{figure}
	\centering
	\includegraphics[width=4.5in]{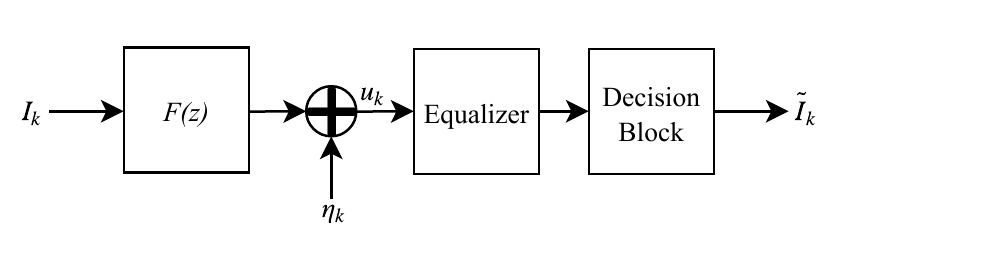}
	\caption{The equivalent discrete time system of the Forney observation model.}
	\label{Equivalent_FOM}
\end{figure}

Figure~\ref{FOM}~\cite{farah:2018} indicates the general system block diagram of the Forney observation model (FOM)~\cite{Forney:1972}.
The received signal in this model is defined as
	\begin{equation}
{r(t)}={\sum_{k}I_kh(t-kT_s)+w(t)},
\end{equation}
where $I_k$ is a symbol drawn from the 16APSK constellation; $T_s$ is the inverse of the symbol rate $R_s$ (symbols/second); $h(t)=g(t)*c(t)$ where $g(t)$ is pulse shape and $c(t)$ is channel impulse response;
${w(t)}$ is a circularly symmetric complex-valued wide-sense stationary normal random process with zero mean and power spectral density $2N_0$ W/Hz.
Maximum likelihood detection applies a matched filter with impulse response $h^*(-t)$ in the receiver. The matched filter output is $y(t)$. The symbol-spaced samples of the matched filter output are $y(kT_s)$.
The FOM uses a discrete time noise whitening filter. The noise whitening filter can be derived by using spectral factorization~\cite{farah:2018}. The output of FOM is
\begin{equation}
u_k=\sum_{n=0}^{L}f_nI_{k-n}+\eta_k,
\label{FOM_EQ}
\end{equation}
where $L+1$ is the length of the channel in the FOM,
$\mathbf{\eta}_k$ is circularly-symmetric complex-valued normal random variables with zero mean and common variance $2N_0$ W/Hz. The system in Figure \ref{FOM} can be re-expressed as the system in Figure~\ref{Equivalent_FOM} by ``using equivalent discrete time channel'' definition.
The output of the Forney observation model is the input to the equalizer. Minimum mean squared error criteria is used in this work to compute the equalizer coefficients, meaning minimum mean squared error equalizer (MMSE). The coefficients of this equalizer can be computed as~\cite{proakis-salehi:2008}
	\begin{equation}
\mathbf{c}=\left[\mathbf{G_f}+\frac{\sigma^2}{E_b}\mathbf{I}\right]^{-1}\boldsymbol{\xi},
\end{equation}
\vspace{5 mm}
where $\mathbf{I}$ is the identity matrix,	\begin{equation}
\mathbf{G_f}=
\begin{bmatrix}
G_f(0)&G_f(-1)&\cdots\\
G_f(1) & G_f(0)& \\
& & \ddots\\
& & & G_f(0)
\end{bmatrix},
\label{G_f_matrix}
\end{equation}
where
\begin{equation}
G_f(k)=\sum_{j=0}^{L} f_{j}f^*_{j-k},
\end{equation}
where $f$ are the channel elements, and
\begin{equation}
\mathbf{\boldsymbol{\xi}}=
\begin{bmatrix}
f_{L} & f_{L-1} & \cdots & f_{0}
\end{bmatrix}'.
\label{eq:xi}
\end{equation}
To come up with the equivalent discrete time observation model different approaches can be used depends on the using channel:
\begin{itemize}
	\item FOM can be generalized to find maximum likelihood combining of the received electrical fields in the cross polarized antenna elements in the receiver, where here two elements are mounted at the directions of $y'$ and $z'$ in the $(x',y',z')$ coordinate system. The block diagram of maximum likelihood of the electrical fields of these two antenna elements is shown in Figure~\ref{fig:ML1}, and its equivalent discrete time observation model is indicated in Figure~\ref{fig:ML1_eq}.
	\begin{figure}
	\centering
\includegraphics[width=7.1in]{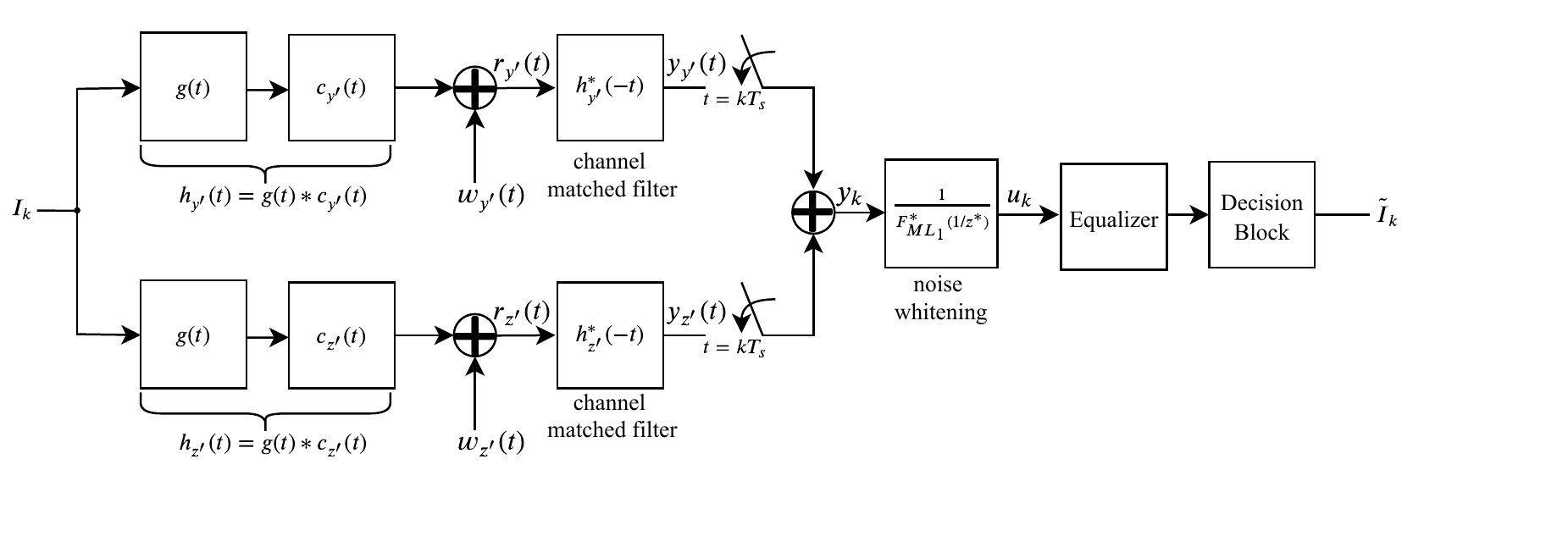}	
	\caption{Maximum likelihood combining of the channels $h_{y'}(t)$ and $h_{z'}(t)$.}
	\label{fig:ML1}
\end{figure}
\begin{figure}
	\centering
	\includegraphics[width=4.5in]{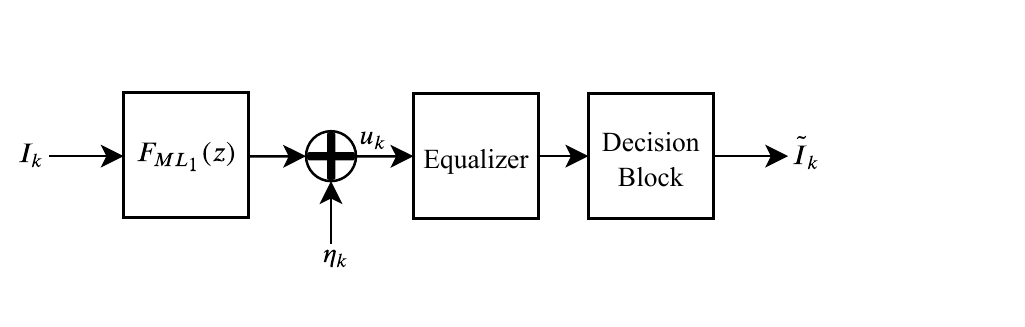}	
	\caption{Equivalent discrete time FOM of maximum likelihood combining of the channels $h_{y'}(t)$ and $h_{z'}(t)$.}
	\label{fig:ML1_eq}
\end{figure}	
	\item FOM can be implemented for the channel $h_{\rm{RHCP}}(t)$, which is defined as
	\begin{equation}
	h_{\text{RHCP}}(t) = \frac{1}{\sqrt{2}}\left(h_{y'}(t) - jh_{z'}(t)\right).
	\label{R}
	\end{equation}
Figure~\ref{fig:RHCP} shows this implementation. Figure~\ref{fig:RHCP_eq} also indicates the equivalent discrete time FOM of the channel $h_{\text{RHCP}}(t)$.
			\begin{figure}
		\centering
		\includegraphics[width=7.1in]{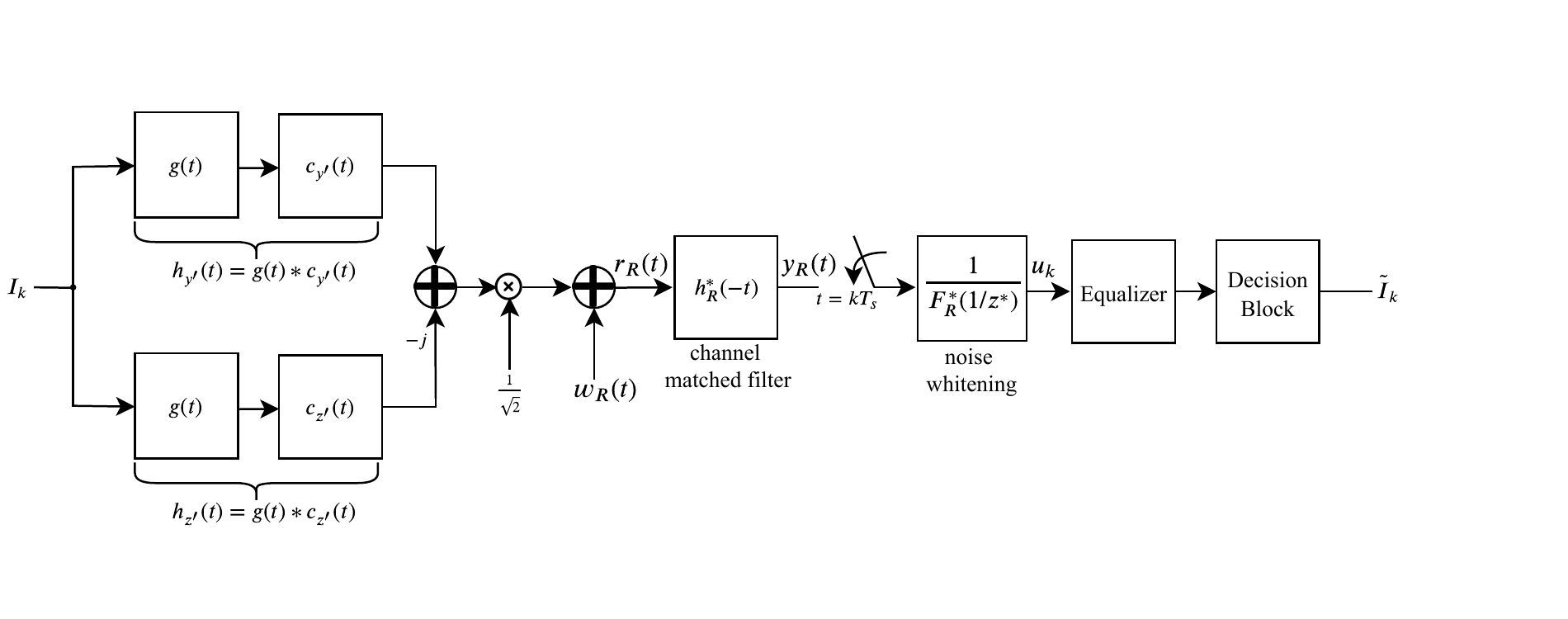}
		\caption{The Forney observation model of the channel $h_{\rm{RHCP}}(t)$.}
		\label{fig:RHCP}
	\end{figure}	
	\begin{figure}
	\centering
	\includegraphics[width=5in]{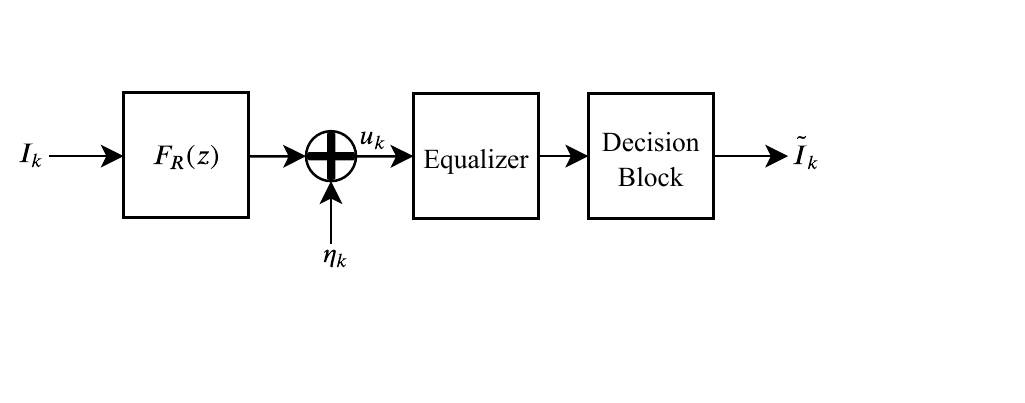}
	\caption{The equivalent discrete time FOM of the channel $h_{\rm{RHCP}}(t)$.}
	\label{fig:RHCP_eq}
\end{figure}
	\item FOM can be implemented for the channel $h_{\rm{LHCP}}(t)$ as shown in Figure~\ref{fig:LHCP}. The $h_{\rm{LHCP}}(t)$ is defined as
	\begin{equation}
	h_{\rm{LHCP}}(t) = \frac{1}{\sqrt{2}}\left(h_{y'}(t) + jh_{z'}(t)\right).
	\label{L} 
	\end{equation}
	Figure~\ref{fig:LHCP_eq} also shows the equivalent discrete time FOM of the channel  $h_{\rm{LHCP}}(t)$.	
	 \begin{figure}
		\centering
		\includegraphics[width=7.1in]{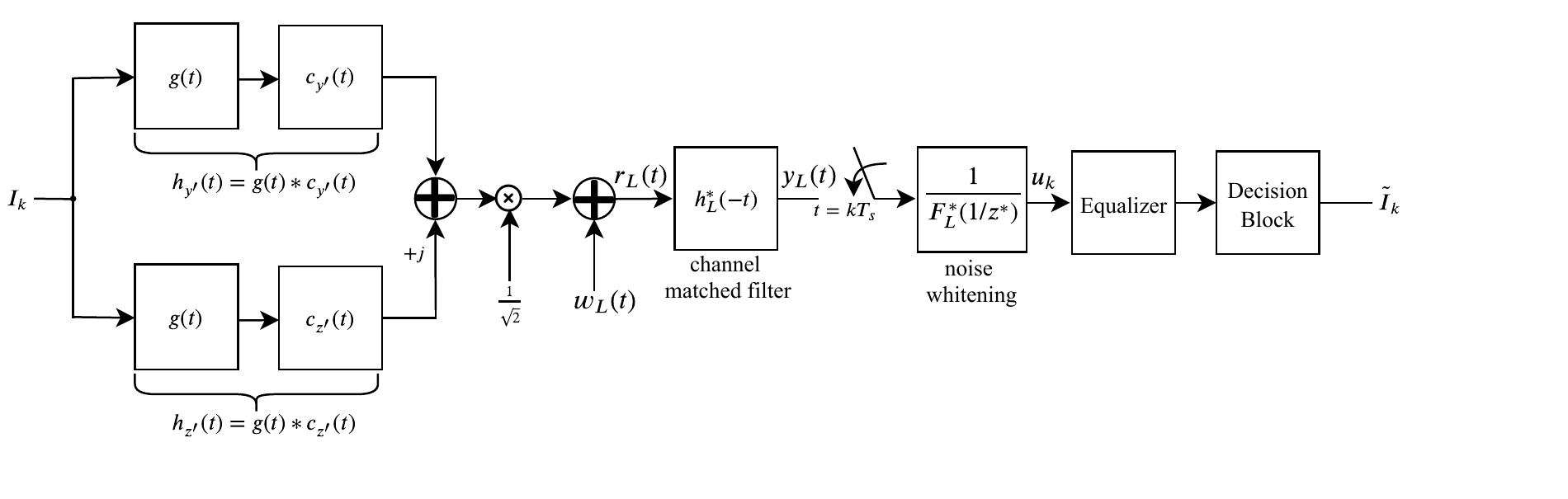}
		\caption{The Forney observation model of the channel $h_{\rm{LHCP}}(t)$.}
		\label{fig:LHCP}
	\end{figure}
	 \begin{figure}
	\centering
	\includegraphics[width=5in]{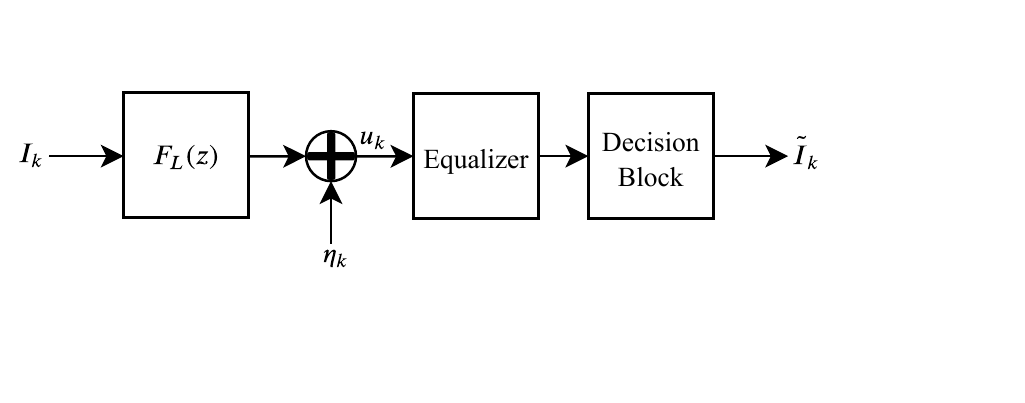}
	\caption{The equivalent discrete time FOM of the channel $h_{\rm{LHCP}}(t)$.}
	\label{fig:LHCP_eq}
\end{figure}
	\item Maximum likelihood combining can be applied to the outputs of ``$90^{\circ}$ hybrid couplers'' that are used in the receiver, meaning the channels $h_{\rm{RHCP}}(t)$ and $h_{\rm{LHCP}}(t)$.	
Figure~\ref{fig:ML2} shows maximum likelihood combining of $h_{\rm{RHCP}}(t)$ and $h_{\rm{LHCP}}(t)$, its equivalent discrete time observation model also is indicated in Figure~\ref{fig:ML2_eq}.
\begin{figure}
	\centering
	\includegraphics[width=7.1in]{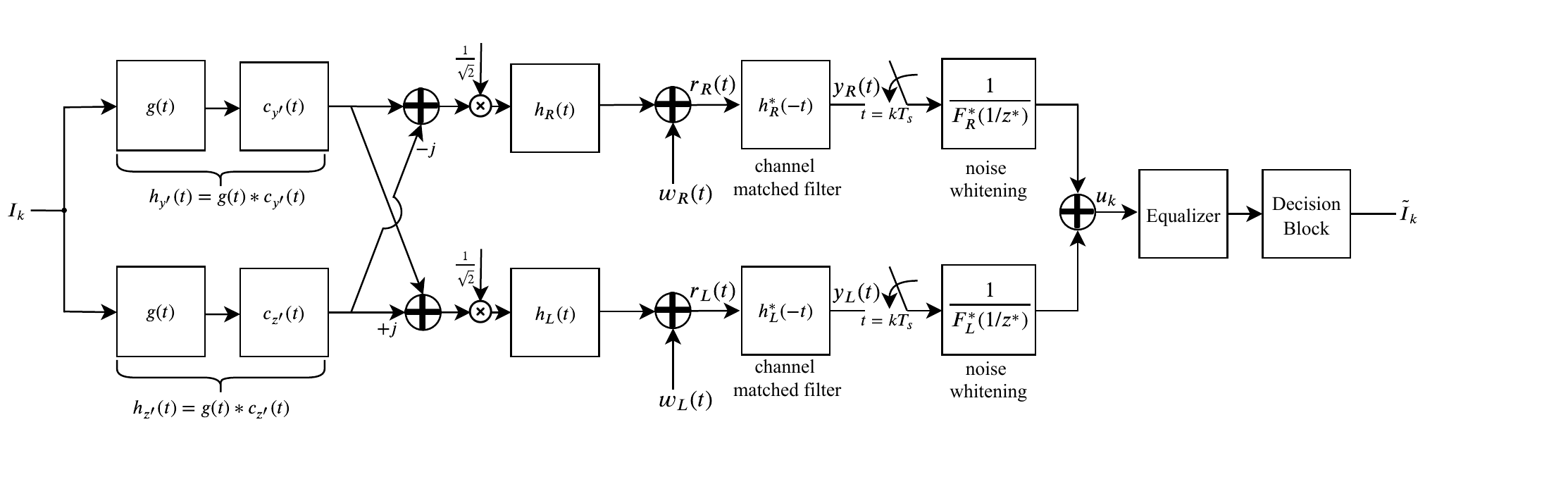}	
	\caption{Maximum likelihood combining of the channels $h_{\rm{RHCP}}(t)$ and $h_{\rm{LHCP}}(t)$.}
	\label{fig:ML2}
\end{figure}	
\begin{figure}
	\centering
	\includegraphics[width=5in]{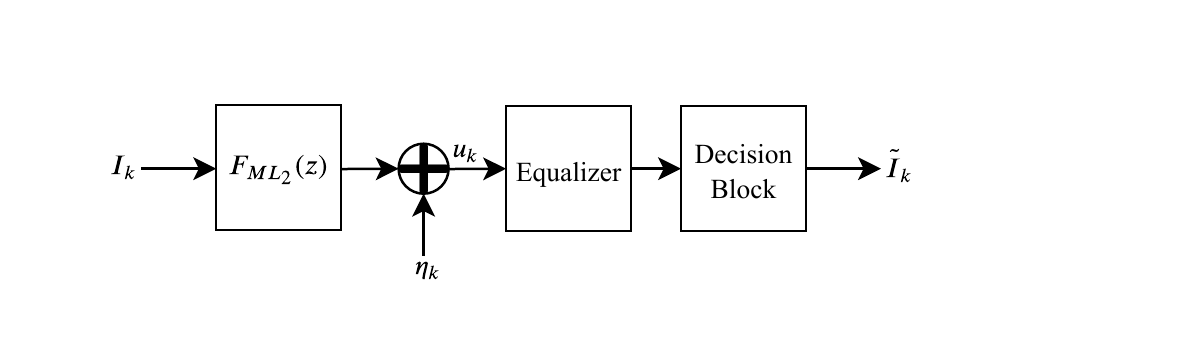}
	\caption{The equivalent discrete time FOM of Maximum likelihood combining of the channels $h_{\rm{RHCP}}(t)$ and $h_{\rm{LHCP}}(t)$.}
	\label{fig:ML2_eq}
\end{figure}
\item The last, but not least implementation that is explored in this paper is combining the channels $h_{y'}(t)$ and $h_{z'}(t)$ before the matched filter as equal gain combining, which is equivalent to the co-phase addition of the outputs of ``$90^{\circ}$ hybrid coupler'', meaning the channels $h_{\rm{RHCP}}(t)$ and $h_{\rm{LHCP}}(t)$. To the best of our knowledge this is one of the common used techniques in the telemetry applications. This approach is shown in Figure~\ref{fig:combiningyz}.
Figure~\ref{fig:combiningyz_eq} indicates the equivalent discrete time FOM of the combining channels of $h_{y'}(t)$ and $h_{z'}(t)$ before the matched filter.
\begin{figure}
	\centering
	\includegraphics[width=7.1in]{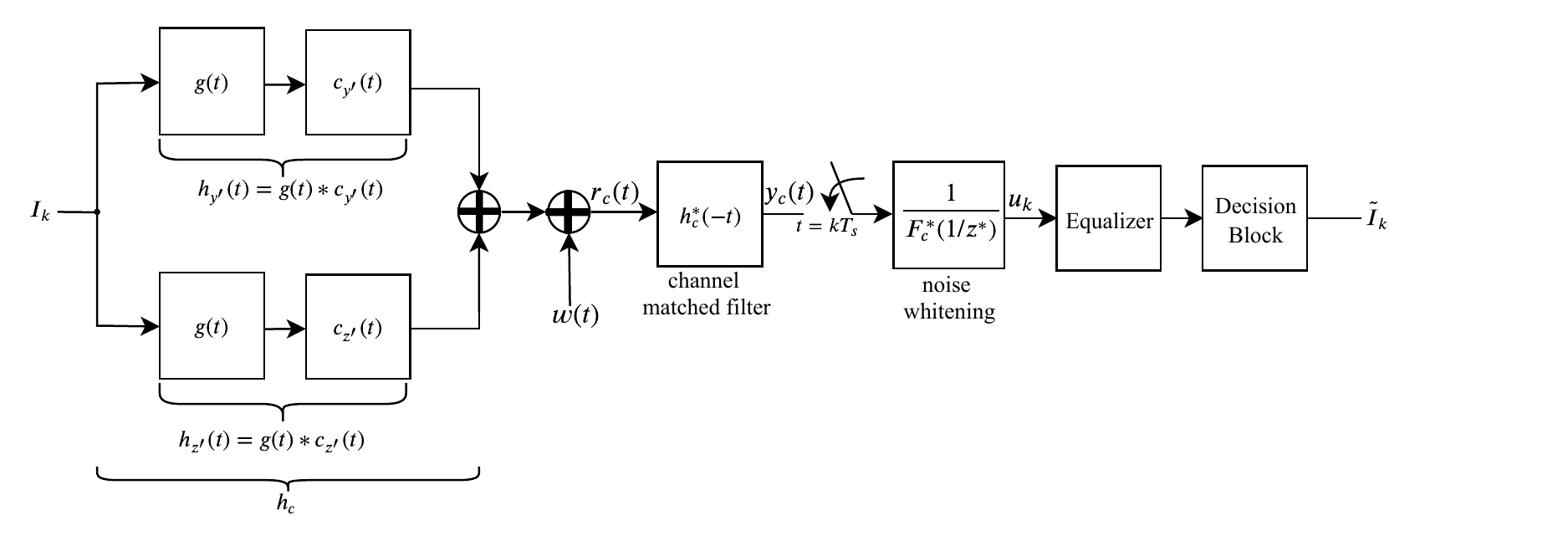}
	\caption{The Forney observation model of the combining of the channels $h_{y'}(t)$ and $h_{z'}(t)$.}
	\label{fig:combiningyz}
\end{figure}
\begin{figure}
	\centering
	\includegraphics[width=5in]{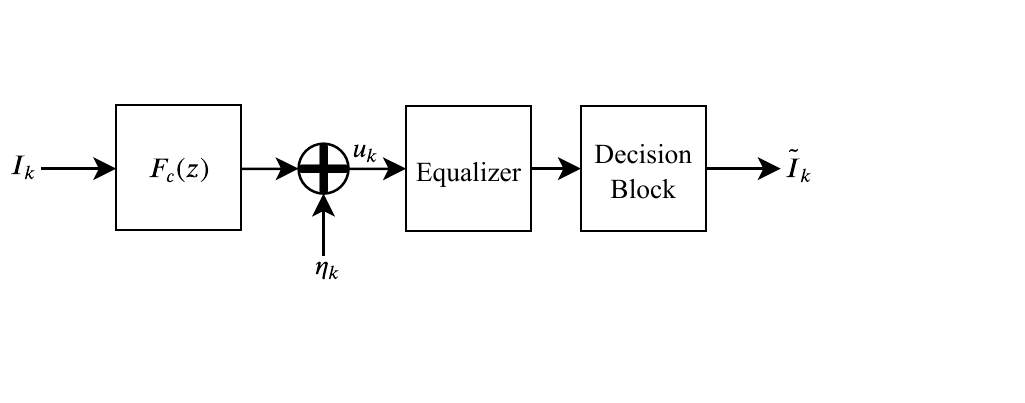}
	\caption{The equivalent discrete time FOM of the combining of the channels of $h_{y'}(t)$ and $h_{z'}(t)$ before the matched filter.}
	\label{fig:combiningyz_eq}
\end{figure}
\end{itemize}
\section{\MakeUppercase{Simulation Results}}
\begin{figure}		
	\centering
	\includegraphics[width=5in]{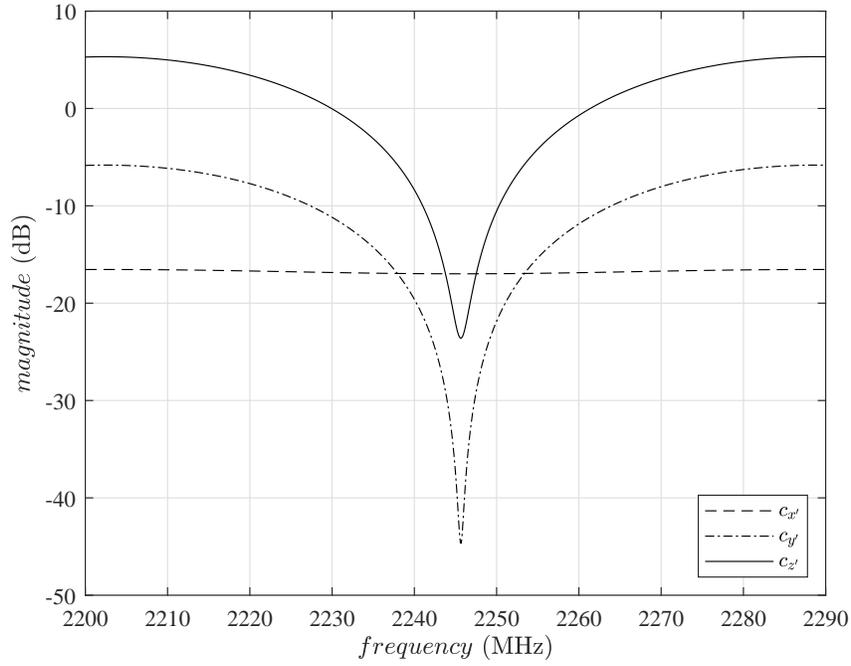}
	\caption{The electrical field components in the receiver, for different directions of $x'$, $y'$ and $z'$ based on frequency while $\theta_{\rm{yaw}}=5^\circ$, $\theta_{\rm{pitch}}=15^\circ$, and $\theta_{\rm{roll}}=10^\circ$. Note the receiver antenna has no element in the $x'$ direction, so this portion of the electrical field will not be detected by the receiver.}
	\label{fig:Efield_y_z}
\end{figure}

The transmitter and receiver antennas geographic location information is indicated in Table~\ref{tab:txrxlocation}.
\begin{table}
	\caption{Transmitter and receiver antennas location.}                                    
	\label{tab:16apsk}
	\begin{center}
		\begin{tabular}{ccccccc}
			\hline\hline
			Antenna & Location & Latitude & Longitude & Altitude (feet AMSL)\\
			\hline\hline
			Transmitter & Cords Road & $35^{\circ}$ $5'$ $0''$ N &$117^{\circ}$ $24'$ $6.73''$ W & 5043 \\ 
			Receiver & Bulding 4795 & $34^{\circ}$ $58'$ $14.63''$ N
			& $117^{\circ}$ $55'$  $52.02''$ W &2710 \\ 
			\hline		
		\end{tabular}
	\end{center}
	\label{tab:txrxlocation}
\end{table}	
The Electrical field representation of the channels in the LOS and NLOS paths can be calculated by math with a noticeable effort, Figure~\ref{fig:Efield_y_z} shows the magnitude of the electrical field representation of the channels in the receiver based on frequency in the $(x',y',z')$ coordinate system while $\theta_{\rm{yaw}}=5^\circ$, $\theta_{\rm{pitch}}=15^\circ$, and $\theta_{\rm{roll}}=10^\circ$.
The $z'$ component has the strongest portion of the electrical field, while  the $y'$ component of the electrical field is weaker, which was expected based on the geometry shown in Figure~\ref{fig:scenario}.

\begin{figure}		
	\centering
	\includegraphics[width=5in]{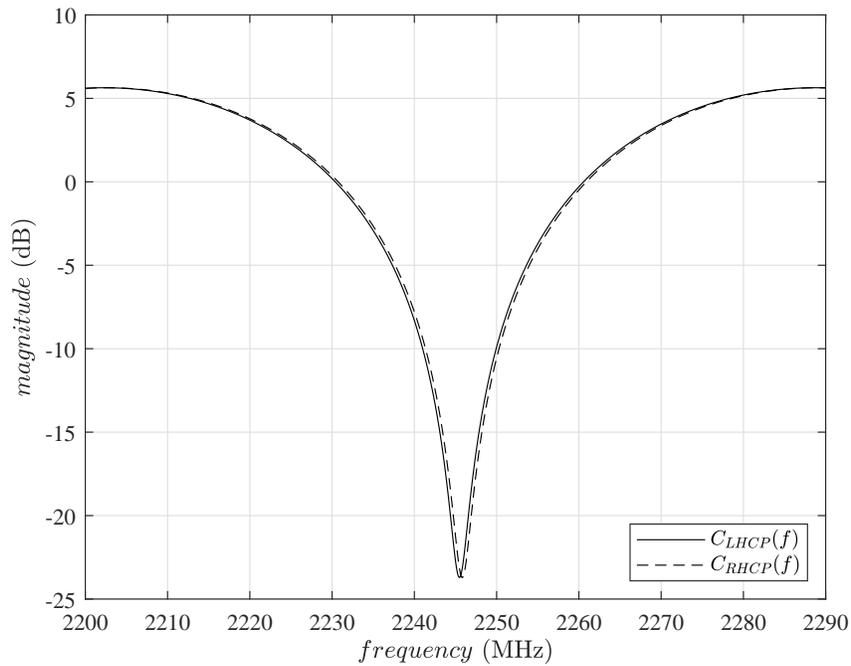}
	\caption{$C_{\text{RHCP}}(f)$ and $C_{\text{LHCP}}(f)$ while $\theta_{\rm{yaw}}=5^\circ$, $\theta_{\rm{pitch}}=15^\circ$, and $\theta_{\rm{roll}}=10^\circ$.}
	\label{fig:Efield_R_L}
\end{figure}

The concept of Equations~\eqref{R} and \eqref{L} can be used to find $C_{\rm{RHCP}}(f)$ and $C_{\rm{LHCP}}(f)$ as shown in Figure~\ref{fig:Efield_R_L}. For pitch rotation equal to zero degree, $C_{\rm{RHCP}}(f)$ and $C_{\rm{LHCP}}(f)$ are identical, but as long as the pitch angle gets bigger, then they will be different by most 1 MHz, that can be useful to achieve gain diversity.
\begin{figure}
	\centering
	\includegraphics[width=6.3in]{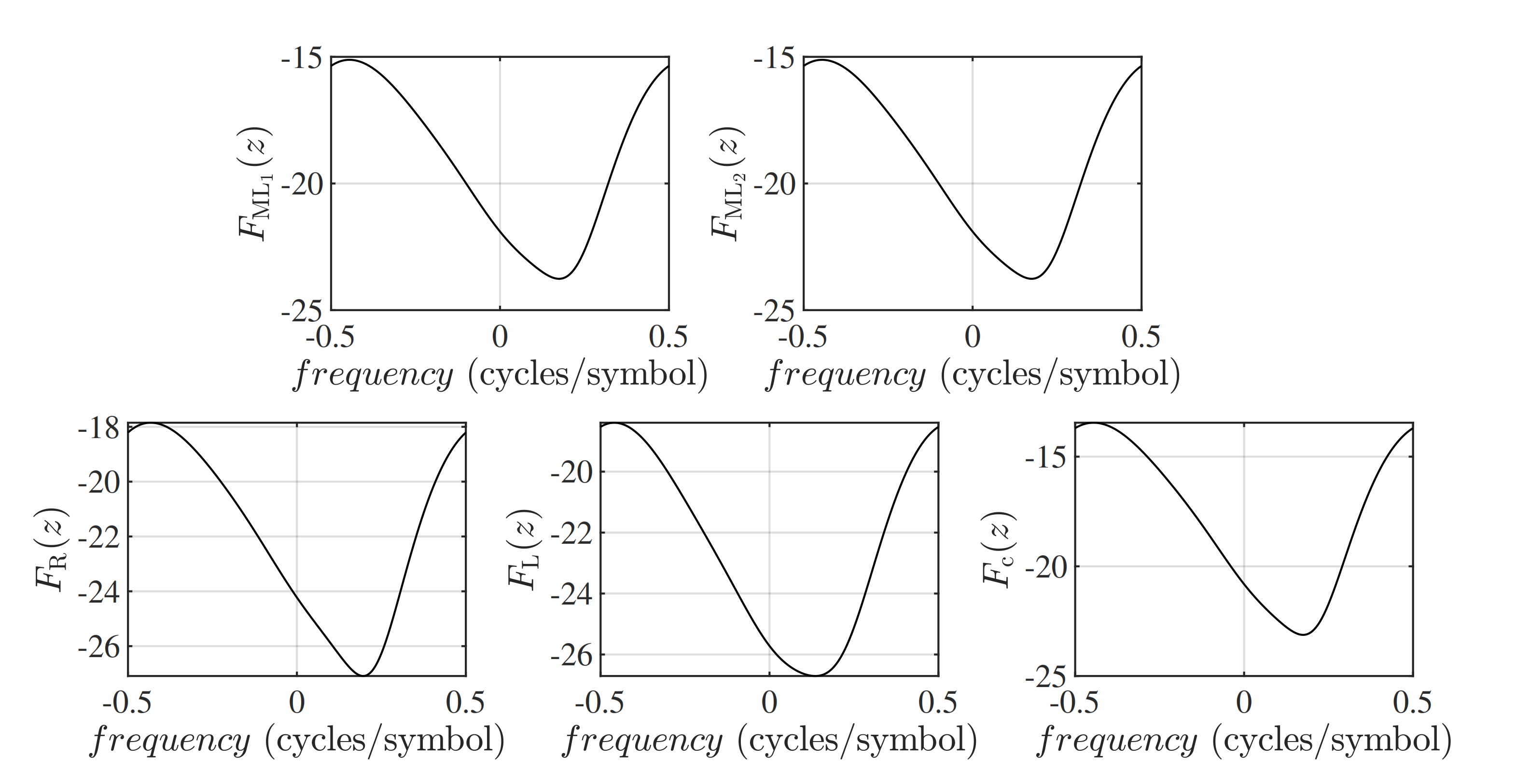}
	\caption{Frequency domain characteristics of the channels used in the simulations.}
	\label{fig:channels}
\end{figure}
Frequency domain characteristics of the channels used in the simulations are outlined in Figure~\ref{fig:channels}.
\begin{figure}
	\centering
	\includegraphics[width=3.3in]{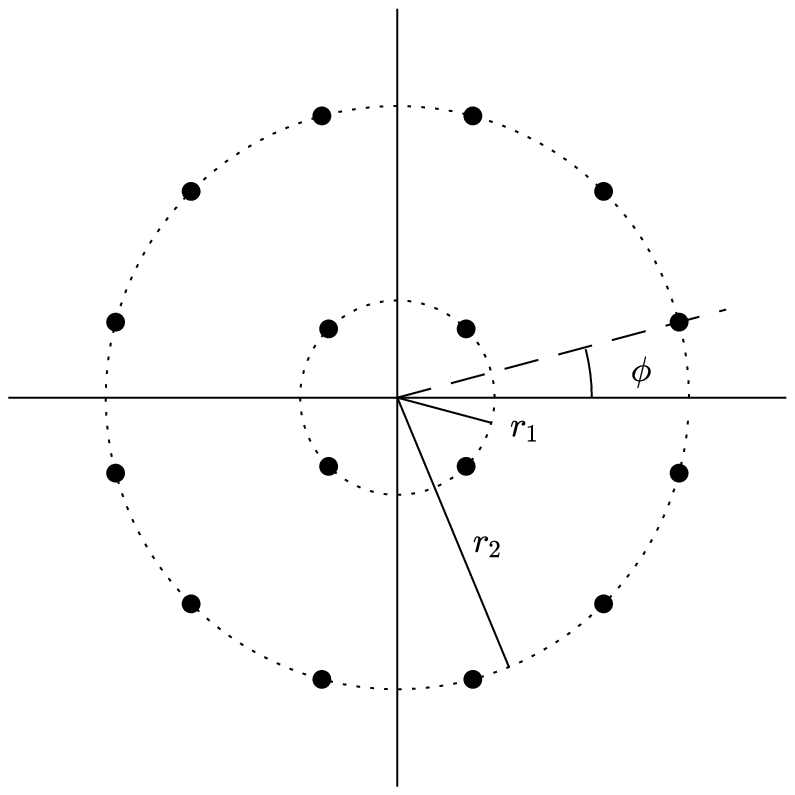}
	\caption{ The 16-APSK constellation from the DVB-S2 standard\cite{ DVB}.}
	\label{fig:16APSK-constellation}
\end{figure}

\begin{figure}
	\centering
	\includegraphics[width=4.8in]{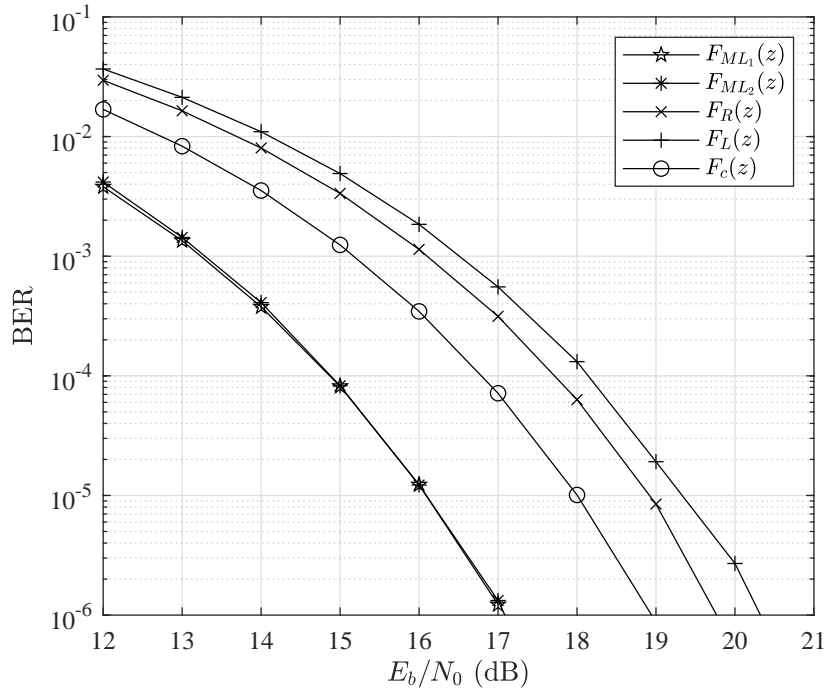}
	\caption{Simulation result for MMSE with FOM, for the equivalent discrete time channels of $F_{ML_1}$, $F_{ML_2}$, $F_R$, $F_L$ and $F_c$.}
	\label{fig:BER}
\end{figure}

BER performance is simulated for 16APSK modulation over all the channels shown in Figure~\ref{fig:channels}.
The 16-APSK constellation is shown in   Figure~\ref{fig:16APSK-constellation}. The constellation is parameterized by the ratio of radii $\gamma = r_2/r_1$ and the phase angle $\phi$. The parameters used in the simulations are those that minimize peak of $Eb/N0$~\cite{Iswcs:2018}: $\gamma = 2.46$ and $\phi = \pi/12$. The reason of choosing 16APSK is having better spectral efficiency than SOQPSK-TG. The pulse shape $g(t)$ is the square-root raised cosine (SRRC) pulse shape with 50\% excess bandwidth~\cite{Michael_Book}. The performance results are shown in Figure~\ref{fig:BER}.
The noticeable observations are as the following:
\begin{itemize}
	\item Maximum likelihood combining of the channels $h_{y'}(t)$ and $h_{y'}(t)$, and the maximum likelihood combining of the channels $h_{\rm{RHCP}}(t)$ and $h_{\rm{LHCP}}(t)$ provide the best performance.
	\item Using the channels $h_{\rm{RHCP}}(t)$ or $h_{\rm{LHCP}}(t)$ each lonely present worse performance among all the simulated channels.
	\item Combining the channels $h_{y'}(t)$ and $h_{z'}(t)$, which is equivalent to combining $h_{\rm{LHCP}}(t)$ and $h_{\rm{RHCP}}(t)$ with ``$90^{\circ}$ hybrid coupler'' does not provide the optimum combining performance, however it outperforms using the channels of $h_{\rm{RHCP}}(t)$ and $h_{\rm{LHCP}}(t)$ separately.
\end{itemize}
\section{\MakeUppercase{Conclusions}}
We present optimal combining of the V and H dipole outputs and optimal combining of the RHCP and LHCP outputs for equalization. We show that the performance of these two optimal combining are almost identical.
We also demonstrate that an equalizer operating on the optimally-combined signal outperforms an equalizer operating on the RHCP signal, LHCP signal, or the combined signal.
\section{ACKNOWLEDGEMENTS}
The funding for this project is managed by the Test Resource
Management Center (TRMC) and funded through the
Spectrum Access R\&D Program under
Contract No. W15QKN-15-9-1004.
\pagebreak
\bibliographystyle{ieeetr}

\end{document}